\documentstyle[aps,prl,twocolumn,floats,epsfig]{revtex}

\newcommand{\be}{\begin{equation}}
\newcommand{\ee}{\end{equation}}
\newcommand{\bea}{\begin{eqnarray}}
\newcommand{\eea}{\end{eqnarray}}
\def\bma{\begin{mathletters}}
\def\ema{\end{mathletters}}

\newcommand{\sdhalf}{\mbox{$\textstyle \sqrt{\frac{D}{2}}$}}
\newcommand{\sdd}{\mbox{$\textstyle \sqrt{\frac{D}{d-1}}$}}

\newcommand{\ket}[1]{ | \, #1  \rangle}
\newcommand{\bra}[1]{ \langle #1 \,  |}
\newcommand{\proj}[1]{\ket{#1}\bra{#1}}

\newcommand{\calu}{\mbox{$\cal U$}}

\newcommand{\mytext}[1]{\mbox{ #1}}

\begin{document}
\draft

\title{Optimal eavesdropping in cryptography with three-dimensional 
quantum states}
\author{ D.~Bru\ss $^1$ and 
C.~Macchiavello$^2$}
\address{
$^1$ Institut f\"ur Theoretische Physik, Universit\"at Hannover, 30167
Hannover, Germany\\
$^2$Dipartimento di Fisica ``A. Volta'' and INFM-Unit\`a 
                 di Pavia,
                 Via Bassi 6, 27100 Pavia, Italy}
\date{Received \today}

\maketitle
\begin{abstract}
We study optimal eavesdropping
 in quantum cryptography with  three-dimensional
systems,
and show that this scheme is more secure 
against symmetric attacks than protocols
using two-dimensional states. We generalize the according eavesdropping
transformation to arbitrary dimensions, and discuss 
 the connection with optimal quantum cloning.
\end{abstract}

\narrowtext

Quantum cryptography, as first suggested by Bennett and Brassard (BB84)
\cite{bb84}, is the experimentally most advanced application
of quantum information processing.
Recently, the use of three-level systems rather than two-level systems
for establishing a secure 
quantum key has been suggested \cite{asher}. The authors study the case
of 4 mutually unbiased bases, i.e. 12 basis states. They consider 
an eavesdropper that uses the most simple strategy, namely 
measuring the state and resending it. For this case they find
that a 3-dimensional system leads to a higher security than a
2-dimensional one.

In order to compare the security of different quantum key distribution
protocols, however, one has to study the most general eavesdropping
attack. This is the aim of our work. Optimal eavesdropping strategies
for the BB84-protocol and the six state protocol have been studied
in \cite{fuchs} and \cite{six,six2}, respectively.

We concentrate our attention to incoherent attacks, namely we assume that
the eavesdropper interacts with a single 3-dimensional quantum system at 
a time. We study the case where the action of the eavesdropper disturbs
all the possible quantum states by
the same amount. Denoting with 
$\{\ket{0},\ket{1},\ket{2}\}$ a basis for the system,
the most general unitary eavesdropping strategy for a set of 3-dimensional
states  can be written as
\bea
\calu \ket{0}\ket{A}& = & \sqrt{1-D}\ket{0}\ket{A_0}+\sdhalf
       \ket{1}\ket{A_1}+\sdhalf\ket{2}
           \ket{A_2}\nonumber \ ,\\
\calu \ket{1}\ket{A}& = & \sdhalf\ket{0}\ket{B_0}+\sqrt{1-D}\ket{1}
           \ket{B_1}+\sdhalf\ket{2}
           \ket{B_2}\nonumber \ ,\\
\calu \ket{2}\ket{A}& = & \sdhalf\ket{0}\ket{C_0}+\sdhalf\ket{1}
           \ket{C_1}+\sqrt{1-D}\ket{2}
           \ket{C_2}\ . \nonumber \\
           &&
\label{u}
\eea
Here $1-D$ is the fidelity of the state that arrives at Bob's site
after Eve's interaction. The disturbance is given by $D$.
We assume the disturbance of the two
basis states that are orthogonal to the original to be equal: this
symmetry is motivated by the fact that the three basis states should be
treated in the same manner. 
The initial state of Eve's  system is called
$\ket{A}$, and her states after interaction are labelled
$\ket{A_0},\ket{B_0},...$ and are normalised. Their dimension
is not fixed.

We have to satisfy unitarity of $\calu$. This leads to
the constraints
\bea
\sqrt{\frac{D(1-D)}{2}}(\bra{B_0}A_0\rangle + \bra{B_1}A_1\rangle )+
   \frac{D}{2}\bra{B_2}A_2\rangle &=&0\ ,\nonumber \\
\sqrt{\frac{D(1-D)}{2}}(\bra{C_2}A_2\rangle + \bra{C_0}A_0\rangle )+
   \frac{D}{2}\bra{C_1}A_1\rangle &=&0\ ,\nonumber \\
\sqrt{\frac{D(1-D)}{2}}(\bra{C_1}B_1\rangle + \bra{C_2}B_2\rangle )+
   \frac{D}{2}\bra{C_0}B_0\rangle &=&0\ .\nonumber \\
   && \,
   \label{uni}
\eea

We consider the cryptographic protocol suggested in Ref. \cite{asher},
where the four mutually unbiased bases are given by 
$\{\ket{0},\ket{1},\ket{2}\}$, and
\bea
&&\{\ket{\alpha} =\frac{1}{\sqrt{3}}(\ket{0}+\ket{1}+\ket{2})\ ,\nonumber \\
&&\ket{\beta} =\frac{1}{\sqrt{3}}(\ket{0}+\omega
\ket{1}+\omega^*\ket{2})\ ,\nonumber \\
&&\ket{\gamma} =\frac{1}{\sqrt{3}}(\ket{0}+\omega^*
\ket{1}+\omega\ket{2})\}\, ;
\eea
\bea
\{\ket{\alpha '} &=&\frac{1}{\sqrt{3}}(\omega
\ket{0}+\ket{1}+\ket{2})\ ,\nonumber \\
\ket{\beta '} &=&\frac{1}{\sqrt{3}}(\ket{0}+\omega
\ket{1}+\ket{2})\ ,\nonumber \\
\ket{\gamma '} &=&\frac{1}{\sqrt{3}}(\ket{0}+\ket{1}
+\omega\ket{2})\}\, ;
\eea
\bea
\{\ket{\alpha ''} &=&\frac{1}{\sqrt{3}}(\omega^*
\ket{0}+\ket{1}+\ket{2})\ ,\nonumber \\
\ket{\beta ''} &=&\frac{1}{\sqrt{3}}(\ket{0}+\omega^*
\ket{1}+\ket{2})\ ,\nonumber \\
\ket{\gamma ''} &=&\frac{1}{\sqrt{3}}(\ket{0}+\ket{1}
+\omega^*\ket{2})\}\;,
\eea
where $\omega = e^{\frac {2\pi i}{3}}$.

We restrict ourselves to the case of symmetric attacks, i.e.
Eve is supposed to introduce an equal disturbance to all 
possible input states written above\footnote{ 
If the noise of the physical device is known to be
symmetric, then Alice and Bob could detect an asymmetric eavesdropper
by checking the error rate in a subset of states. Otherwise, the
trade-off between Eve's information and the signal key is more
complicated to handle.}.
We can then
directly compare the security to 
the six state scheme 
for qubits, where only
symmetric attacks have  been studied.
  By imposing that the
disturbance $D=1-\mytext{Tr}(\proj{\psi_i}\varrho^{out}_B)$, 
where $\varrho^{out}_B$ is the reduced density operator of the
state sent on to Bob, takes the same value for all 12 possible
input states $\ket{\psi_i}$,
we  derive the following relations that involve the scalar products
of Eve's output states:

\bea
&&\sqrt{2D(1-D)}(\bra{A_1}A_0\rangle + \bra{B_1}B_0\rangle
+\bra{C_2}B_0\rangle + \bra{A_1}C_2\rangle)\nonumber\\
&&\ \ \ +D (\bra{C_1}C_0\rangle+3\bra{B_0}A_1\rangle)=0\ , 
\label{cond1} \\
&&\sqrt{2D(1-D)}(\bra{B_1}C_0\rangle + \bra{A_2}B_1\rangle
+\bra{A_2}A_0\rangle + \bra{C_2}C_0\rangle)\nonumber\\
&&\ \ \  +D (\bra{B_2}B_0\rangle+3\bra{C_0}A_2\rangle)=0\ ,
\label{cond2}\\
&&\sqrt{2D(1-D)}(\bra{B_2}B_1\rangle + \bra{C_2}C_1\rangle
+\bra{B_2}A_0\rangle + \bra{A_0}C_1\rangle)\nonumber\\
&&\ \ \  +D (\bra{A_2}A_1\rangle+3\bra{C_1}B_2\rangle)=0\ ,
\label{cond3}\\
&&\bra{A_1}C_0\rangle + \bra{A_2}B_0\rangle
+\bra{B_0}C_1\rangle \nonumber\\
&&\ \ \  + \bra{B_2}A_1\rangle+\bra{C_1}A_2\rangle+\bra{C_0}B_2\rangle=0\ .
\label{cond4}
\eea

Note that both real and imaginary part of these expressions have to
vanish. Writing the disturbance 
introduced through the eavesdropping transformation (\ref{u})
as a function of the scalar products
of Eve's states, and taking into account unitarity (\ref{uni})
and the conditions 
(\ref{cond1})-(\ref{cond4}), we find the following simple form:

\be
D=2\frac{1-S}{3-2 S}\;, 
\label{d}
\ee
where $S={\mbox{Re}}[\bra{A_0}B_1\rangle+\bra{B_1}C_2\rangle+
\bra{C_2}A_0\rangle]/3$.
Notice that in the expression for the disturbance only the scalar 
products among the eavesdropper's states $\ket{A_0},\ket{B_1}$ 
and $\ket{C_2}$ appear, while all the others do not contribute.

We will now derive the optimal eavesdropping transformation for a fixed value
$D$ of the disturbance, namely we maximise the mutual information 
$I_{AE}$ between Alice and Eve. 
(This is a standard figure of merit for the description of the efficiency of
an eavesdropping attack \cite{fuchs}.)
As mentioned above,  the disturbance 
introduced by Eve is independent of the scalar products of 
her states, apart from the ones involving $\ket{A_0},\ket{B_1}$ 
and $\ket{C_2}$. Therefore,
for any value of $D$,  Eve is free to choose those states on which
$D$ does not depend in such a way that she
retrieves the maximal information. The optimal  choice is to take 
all of these states orthogonal to each other, because in this case
Eve can infer the original state sent by Alice in an unambiguous way
from her measured state.

We will now consider only the scalar products that appear in $S$ and 
choose them such that the mutual information is maximised for fixed $S$,
i.e. for a given disturbance $D$.
We introduce the general parametrisation for the normalised
auxiliary states,
 \bea
 \ket{A_0}&=&x_A\ket{\bar 0}+y_A\ket{\bar 1}+z_A\ket{\bar 2}\ ,\nonumber \\
  \ket{B_1}&=&x_B\ket{\bar 0}+y_B\ket{\bar 1}+z_B\ket{\bar 2}\ ,\nonumber \\
  \ket{C_2}&=&x_C\ket{\bar 0}+y_C\ket{\bar 1}+z_C\ket{\bar 2}\ ,
  \label{abc}
 \eea 
 where $\{\ket{\bar 0},\ket{\bar 1},\ket{\bar 2}\}$ is an orthonormal
 basis which is orthogonal to all the other auxiliary states.
 In order to treat the basis states $\ket{0},\ket{1},\ket{2}$
 in the same way, we require that the overlaps 
of these three states are equal. We choose $x_A=y_B=z_C=x$, 
while all other coefficients are equal. Without loss of generality
we can take the coefficients to be real.

 With this strategy we find the optimal
  mutual information between
 Alice and Eve to be
 \bea
 I_{AE} &=& 1+(1-D)[f(D)\log_3 f(D) \nonumber \\
     && \ \ \ \ \ \ \ \ \ \ \ \ 
     +(1-f(D))\log_3\frac{1-f(D)}{2}]\ ,
     \label{iae}
 \eea
 where $f(D)$ is given by
 \be
 f(D)=\frac{3-2D+2\sqrt{2}\sqrt{D(3-4D)}}{9(1-D)}\ .
 \ee
 The relation between $x$ and $D$ is $x^2=f(D)$.
 Inserting this into equations (\ref{abc}) leads, together with the
 ansatz (\ref{u}) and a straightforward choice of the ancilla states,
 to the explicit form of the optimal transformation. 
 Eve needs to employ
 two three-level systems for the optimal attack.

 The information for Bob decreases with increasing  disturbance:
 \be
 I_{AB}=1+(1-D)\log_3(1-D)+D\log_3\frac{D}{2}\ .
 \label{iab}
 \ee
 
  Note that we renormalized the functions given in (\ref{iae})
  and (\ref{iab}), as in \cite{asher},
   in order to be able to directly relate the
  values to the 2-dimensional case.
  
  We will now compare the security of the 3-dimensional
  scenario as described above with the most secure
  2-dimensional scheme, that employs six states
  (i.e. three mutually unbiased bases) \cite{six,six2}. The
  according information curves of both protocols are
  shown in figure \ref{Fig.1}.

\begin{figure}[ht]
\vspace*{-1.5cm}
\begin{picture}(150,150)
\put(5,5){\epsfxsize=280pt\epsffile[23 146 546 590 ]{info.ps}}
\end{picture}
\vspace*{2.cm}
\caption[]{\small Mutual information for
Alice/Bob and Alice/Eve as a function of the disturbance, for
2-dimensional and 3-dimensional quantum states.}
\label{Fig.1}
\end{figure}

 We find that the 3-dimensional protocol is more
 secure in two respects: first, the information curves for Bob and
 Eve intersect at a higher disturbance $D_c$ than for the 2-dimensional
 case,
 namely $D_{c,3}=0.227$, while $D_{c,2}=0.156$.
  In other words, Eve has to introduce {\em more} noise in order
 to gain the same information as Bob. 
 In general, for disturbances $D<D_c$, a key distribution protocol
 can be considered secure, because $I_{AB}>I_{AE}$ \cite{fuchs}.
 Therefore, the 3-dimensional protocol is secure up to higher 
 disturbances.
 Second, for a fixed 
 disturbance $D<D_c$, Bob gets more and Eve less information than in
 the 2-dimensional case. The price that has to be payed for higher
 security is a lower efficiency: the basis for Bob matches the one
 of Alice in fewer cases than for two dimensions, as the number
 of bases is increased.
 
 Notice that our derivation of the optimal eavesdropping
 transformation relies on equations  (\ref{cond1})-(\ref{cond4})
 which guarantee that all the possible input states
 are disturbed in the same way. If we reduce the number of
 bases, not all of these conditions will be necessary, thus
 leading to a less simple structure of $D$ than the one given
 in (\ref{d}). This would allow a different general form of the
 optimal eavesdropping transformation, and a higher curve
 for $I_{AE}$. The analogous behaviour was shown for
 the 2-dimensional case in \cite{six,six2}, where  the
 six-state protocol and the BB84 scheme were compared.

Generalising the ansatz given in (\ref{u}) and
the  structure of the ancilla states 
as in (\ref{abc}) to higher dimensions,
we find a lower bound on the eavesdropper's information for
quantum cryptography with $d$-dimensional systems.
The general ansatz is then
\bea
\calu \ket{0}\ket{A}& = & \sqrt{1-D}\ket{0}\ket{A_0}+\sdd
       \ket{1}\ket{A_1}+... \nonumber \ ,\\
\calu \ket{1}\ket{A}& = & \sdd\ket{0}\ket{B_0}+\sqrt{1-D}\ket{1}
           \ket{B_1}+ ...\nonumber \ ,\\
    &\vdots & \nonumber \\
\calu \ket{d-1}\ket{A}& = & \sdd\ket{0}\ket{Z_0}+\sdd\ket{1}
           \ket{Z_1}+ ...\ .\nonumber \\
    \       & \, &
\label{ug}
\eea
(The alphabet denoting Eve's states is supposed to contain $d$ letters.)
The according generalized formula for the disturbance
as a function of the scalar products 
is
\be
D=\frac{(d-1)(1-S)}{d- S(d-1)}\;, 
\label{dg}
\ee
where $S$ is now the real part of the average of all possible scalar products
between $\ket{A_0},\ket{B_1},...$.
The function $f$ is then given by
\be
 f_d(D)=\frac{d-2D+\sqrt{(d-2D)^2-d^2(1-2D)^2}}{d^2(1-D)}\ .
\ee
In figure \ref{Fig.2} we plot Eve's corresponding  information
\bea
 I_{AE,d} &=& 1+(1-D)[f_d(D)\log_d f_d(D) \nonumber \\
     && \ \ \ \ \ \ \ \ \ \ \ \ 
     +(1-f_d(D))\log_d\frac{1-f_d(D)}{d-1}]\ ,
     \label{iaed}
 \eea
 as a function of the dimension $d$ for a fixed value of the
 disturbance $D$. We conjecture that this mutual information
 is optimal when employing the maximal number of mutually unbiased
 bases for a given dimension \cite{muba}.
 
\begin{figure}[ht]
\vspace*{-1.5cm}
\begin{picture}(150,150)
\put(5,5){\epsfxsize=280pt\epsffile[23 146 546 590 ]{dinfo1.ps}}
\end{picture}
\vspace*{2.cm}
\caption[]{\small Mutual information between 
 Eve and Alice as a function of the dimension, for
$D=0.1$.}
\label{Fig.2}
\end{figure}

Finally, we discuss the connection between optimal eavesdropping
strategies and optimal cloning transformations. 
The information that Eve can gain is restricted by the laws
of quantum mechanics, namely the no-cloning theorem \cite{wootters}. 
Let us point out, however,  that there is, in general, no direct 
connection between limits on the cloning fidelity for a given 
$d$-dimensional state, and the intersection of the information
curves of Bob and Eve. The reason is that approximate cloning 
transformations \cite{cloning}
are only a subset of our family of transformations
  \calu\  given in eq. (\ref{u}), because an additional symmetry between
  the first of Eve's states and Bob's state is required for
  cloning. Indeed, if Eve would read only the first of her two
  states, the disturbance for the intersection between the two
  resulting information curves would correspond to the fidelity of the optimal
  cloner. Reading both states increases her information. Therefore,
  the knowledge of cloning transformations for $d$-dimensional systems
  \cite{dclone}
  allows only to find  a lower bound on Eve's information at a 
  given disturbance.
  
In summary, we have found a remarkable feature of higher-dimensional
quantum systems: we have proven analytically 
for dimension $d=3$ that the most general
symmetric attack of an eavesdropper gives her less information than 
in the case of qubits. 
Therefore a three-dimensional scheme offers
higher security than two-dimensional systems. 
We generalised the upper limit for Eve's information 
$I_{AE}$ from $d=3$
to higher dimensions: this  limit decreases with the dimension,
and numerically we find that it
reaches $I_{AE}=D$ in the limit $d\rightarrow \infty$.
As quantum cryptography is the most advanced 
technology in quantum information, and  security issues play
a fundamental role in any study of cryptography,
  it is important to discuss quantitative 
properties of the security in quantum key distribution: here quantity 
becomes quality.
 
While completing this manuscript we learnt about related
work by  M. Bourennane et al \cite{mb}.

We wish to thank Maciej Lewenstein for  discussions.
This work has been supported  by  DFG 
(Schwer\-punkt  ``Quan\-ten\-informationsverarbeitung"), 
the ESF-Programme PESC,  and the EU IST-Programme
EQUIP.


\begin{thebibliography}{99}
\bibitem{bb84} C.~H.~Bennett and G.~Brassard, in {\em Proceedings of the IEEE
International Conference on Computers, Systems, and Signal Processing, Bangalore,
India} (IEEE, New York, 1984), pp. 175-179.
\bibitem{asher} H.~Bechmann-Pasquinucci and A.~Peres, quant-ph/0001083.

\bibitem{fuchs}  C.~Fuchs, N.~Gisin, R.~Griffiths, 
C.-S.~Niu and A.~Peres, Phys. Rev. A {\bf 56}, 1163 (1997).

\bibitem{six}  D. Bru{\ss}, Phys. Rev. Lett. {\bf 81}, 3018 (1998).

\bibitem{six2}  H. Bechmann-Pasquinucci and N. Gisin, 
           Phys. Rev. A {\bf 59}, 4238 (1999).
           
\bibitem{muba}  S. Bandyopadhyay, P.  Boykin, 
 V. Roychowdhury and F. Vatan, quant-ph/0103162.
 
 \bibitem{wootters}  W.K.~Wootters and W.H.~Zurek, Nature {\bf 299}, 802
(1982).

\bibitem{cloning} V. Bu\v{z}ek and M.~Hillery, Phys.\ Rev.\ A {\bf 54},
1844 (1996);
N.~Gisin and S.~Massar, Phys.~Rev.~Lett. {\bf 79}, 2153
(1997);
D. Bru\ss , D. P. DiVincenzo, A. Ekert, C. A. Fuchs, C. Macchiavello
and  J. A. Smolin, Phys.~Rev. A{\bf 57}, 2368 (1998);
R.~Werner, Phys.~Rev. A{\bf 58}, 1827 (1998).

\bibitem{dclone} S.~Albeverio and  S.-M.~Fei,
           Eur. Phys. J. B {\bf 14}, 669 (2000).
           
\bibitem{mb}  M. Bourennane, A. Karlsson, G. Bj\"ork, 
N. Gisin and N. Cerf, quant-ph/0106049;
N. Cerf, M. Bourennane, A. Karlsson and N. Gisin,
quant-ph/0107130.

\end{thebibliography}
\end{document}